\documentstyle[12pt]{article}

\def\be{\begin{equation}}
\def\ee{\end{equation}}
\def\bea{\begin{eqnarray}}
\def\eea{\end{eqnarray}}
\def\a{\alpha}
\def\b{\beta}

\def\l{\lambda}

\def\z{\zeta}

\def\f{\phi}
\def\F{\phi}

\newcommand{\non}{\nonumber}
\begin{document}

\begin{titlepage}
\begin{flushright}
KL --  TH -- 94/23 \\
hep-th/9410040
\end{flushright}

\vspace{1cm}

\begin{center}
{\LARGE CLASSICAL NON-ABELIAN SOLITONS }     \\ \vspace{0.6cm}
\end{center}

\vspace{.5cm}

\begin{center}
{\large V. F. M\"uller}\footnote{E--mail: \ vfm@physik.uni-kl.de} \\
{\it Fachbereich Physik  \\Universit\"at Kaiserslautern \\67653 Kaiserslautern,
Germany}
\end{center}
\vskip .2 cm

\vskip 1.5 cm
\vskip 1.5 cm

\begin{abstract}
In two space-time dimensions a class of classical multicomponent scalar field
theories with discrete, in general non-Abelian global symmetry is considered.
The corresponding soliton solutions are given for the cases of 2, 3,  and
4 components.
\end{abstract}

\end{titlepage}
\setcounter{page}{1}

Recently renewed interest in solitons has arisen in connection with exceptional
statistics occuring in low-dimensional quantum field theory. The
nonperturbative
approach to quantum solitons \cite{H, MS, MSS, M, FM}, based on the notion of
a disorder variable \cite{KC, FK}, does not make use of the well-known
semiclassical quantisation procedure around classical soliton solutions
\cite{RS}.
In a recent article \cite{VFM} the author introduced multicomponent scalar
field
models, treated nonperturbatively on a Euclidean space-time lattice. The
exponentially decaying disorder correlation functions are connected with
soliton
fields showing non-Abelian braid group statistics. It is the aim of this note
to
present the corresponding classical soliton solutions, which do not seem to
have
appeared in the literature.

In the two-dimensional Minkowski space with its space-time points denoted by
$(t, x)$ the models considered in \cite{VFM} look as follows. A set of $n$ real
scalar fields $\{\F_k(t,x)\}$, $k = 1, \ldots, n$ , is introduced with the
classical
Lagrangian density functional
\be
L(\{\F_k(t,x)\}) = \frac{1}{2} \sum_{k=1}^{n} \{( \partial_{t} \F_k(t,x))^2
- ( \partial_{x} \F_k(t,x))^2 \} - v (\{\F_k(t,x)\})
\label{1}
\ee
The interaction $v$ is a fourth order polynomial, with $\l, \xi, b \in {\bf
R_{+}}$,
\bea
v(\{\F_k\}) &=& \l \{(|\F|^2 - \xi)^2 + 4b\sum_{1 \leq k < l \leq n} (\F_k
\F_l)^2
\label{2} \\
|\F|^2        &=& \sum_{k=1}^{n} \F_k^2.
\label{3}
\eea
The Lagrangian density (\ref{1}) for given $n$ is invariant under a
corresponding
global discrete group $G$ transforming the fields. Moreover, the interaction
(\ref{2}) has a degenerate absolute minimum on which already a subgroup
$G_0 \subset G$ acts transitively. The cases $n= 2, 3, 4$ are treated.

\noindent
$i$) $n=2$: We introduce the complex field $\F = \F_1 + i\F_2$ , then $G_0$ is
the
abelian group $Z_4$ and $v$ has a fourfold degenerate absolute minimum at
$\F = \z_m$, $m = 1, \ldots 4$
\be
\z_1 = -\z_3 = \sqrt{\xi},  \quad  \z_2 = -\z_4 = i \sqrt{\xi}
\label{4}
\ee

\noindent
$ii$) $n=3$: We view $\f = (\f_1, \f_2, \f_3 ) \in {\bf R^3}$ as a real vector
field, then
$G_0$ is the octahedron group acting as a subgroup of the standard
representation of $O(3)$ and $v$ has a sixfold degenerate absolute minimum at
$\f = \z_m$, $m = 1, \ldots, 6$
\be
\z_1 = -\z_3 = (\sqrt{\xi}, 0, 0),  \quad \z_2 = -\z_4 = (0, \sqrt{\xi}, 0),
\quad
\z_5 = -\z_6 = (0, 0, \sqrt{\xi})
\label{5}
\ee

\noindent
$iii$) $n=4$: We introduce the complex vector field $\f = (\f_1 +i\f_3, \f_2
+i\f_4)
\in {\bf C^2}$ and have $G_0 = \{\pm \sigma_0, \pm i\sigma_1, \pm i\sigma_2,
\pm i\sigma_3 \} \subset SU(2)$, with the unit matrix $\sigma_0$ and the Pauli
matrices
$\sigma_k$, $k = 1, 2, 3$. The interaction $v$ has an eightfold degenerate
absolute minimum at $\f = \z_m$, $m = 1, \ldots, 8$
\bea
\z_1 &=& -\z_5 =- i \z_3 =i\z_7 = (\sqrt{\xi}, 0),   \non  \\
\z_2 &=& -\z_6 = -i\z_4 = i\z_8 = (0, \sqrt{\xi})
\label{6}
\eea

In the following we restrict to the distinguished value $b=1$ in (\ref{2}).
Then
(\ref{1}) implies the classical field equations
\be
\{ (\partial_t)^2 - (\partial_x)^2 \} \f_{k} + 4\l \f_{k} \{ |\f|^2 - \xi +
2 ( |\f|^2  - \f_k^2) \} =0,
\label{7}
\ee
\bea
k = 1, \ldots, n,  \non
\eea
and the Hamiltonian functional
\be
H( \{ \pi_k, \f_k\} ) = \int_{-\infty}^{\infty}dx \{ \frac{1}{2} \sum_{k=1}^{n}
[ (\pi_k(t,x))^2 + (\partial_x \f_k(t,x))^2 ] + v( \{\f_k(t,x)\}) \} \geq  0
\label{8}
\ee
where $\pi_k$ is the momentum canonically conjugate to $\f_k$.

In the cases $n = 2, 3, 4$ considered there are  two classes of stationary
classical solutions $\f(t,x) = \varphi(x)$ of (\ref{7}) having finite energy:
$\a )$ the obvious vacuum solutions $\varphi(x) \equiv \z_m$,
$\b)$ soliton solutions $\varphi(x)$ satisfying $\varphi(+\infty) = \z_l$,
$\varphi(- \infty) = \z_m$, $l \neq m$, to be presented. The conserved
topological
current
\be
j_{k}^{\mu}(t,x) = \varepsilon^{\mu \nu} \partial_{\nu} \f_{k} (t,x)
\label{9}
\ee
implies for a given  solution $\varphi(x)$ the ``topological charge''
$\varphi(+\infty) - \varphi(- \infty)$, which is a vector in ${\bf C}$, ${\bf
R^3}$
and ${\bf C^2}$
respectively, for the cases considered.

Due to the symmetry group $G_0$ the individual minima in the degenerate
absolute minimum are physically equivalent. Hence it suffices to list only
solutions having the boundary value  $\varphi(+\infty) = \z_1$, with
$\z_1$ selected arbitrarily. We label the different soliton types by the
respective maps $T$ defined by
\be
\varphi(- \infty) = T\varphi(+\infty).
\label{10}
\ee
Then the  topological charge can be written as
\be
\varphi(+\infty) - \varphi(- \infty) = (1 -T)\varphi(+\infty).
\label{11}
\ee
If $\varphi(+\infty)$ has a stabilizer group $H \in G_0$, as in $ii)$, $T$ is
any
representative of the coset decomposition of $G_0$ with respect to $H$.

Listing the soliton solutions the shorthand
\be
y = (x - s) \sqrt{2 \l \xi}
\label{12}
\ee
with $s \in {\bf R}$ is used.

\noindent
$i$)
\bea
\varphi(x) \in {\bf C}  &;&  \varphi(-\infty) = e^{i \a} \z_1
\label{13}
\eea
There are 3 different soliton types
\bea
e^{i \a} = -1     &:&     \varphi(x) = \sqrt{\xi} \tanh y
\label{14} \\
e^{i \a} = \pm i &:&     \varphi(x) = \frac{1}{2}  \sqrt{\xi} \{ 1 + \tanh y
\pm i (1 - \tanh y) \}
\label{15}
\eea

\noindent
$ii$)
\bea
\varphi(x) \in {\bf R^3}  &;&
               \varphi_{k}(-\infty) = \sum_{l=1}^3 D_{kl}(\z_1)_l, \quad k = 1,
2, 3.
\label{16}
\eea
There are  5 different soliton types. Denoting by $D^{(l )}(\a)$ the rotation
around the $l$ - axis with the rotation angle $\a$ we find
\bea
D = D^{(3)} (\pi)                       &:&   \varphi(x) = \sqrt{\xi} ( \tanh
y, 0, 0)
\label{17}  \\
D = D^{(3)} (\pm \frac{\pi}{2})   &:&   \varphi(x) = \frac{1}{2}\sqrt{\xi}
                              (1 + \tanh y, \pm (1 - \tanh y), 0)         \\
\label{18}
D = D^{(2)} (\pm \frac{\pi}{2})   &:&   \varphi(x) = \frac{1}{2}\sqrt{\xi}
                             (1 + \tanh y , 0, \mp (1 - \tanh y))
\label{19}
\eea

\noindent
$iii$)
\bea
\varphi(x) \in {\bf C^2} &;&
\varphi_{k}(-\infty) = \sum_{l=1}^2 u_{kl}(\z_1)_l, \quad k = 1, 2.
\label{20}
\eea
There are 7 different soliton types
\bea
u = - \sigma_0        &:&  \varphi(x) =  \sqrt{\xi} ( \tanh y, 0)
\label{21}   \\
u = \pm i \sigma_1  &:&  \varphi(x) = \frac{1}{2}\sqrt{\xi}
                              (1 + \tanh y, \pm i (1 - \tanh y))
\label{22}  \\
u = \pm i \sigma_2  &:&   \varphi(x) = \frac{1}{2}\sqrt{\xi}
                             (1 + \tanh y ,  \mp (1 - \tanh y))
\label{23} \\
u = \pm i \sigma_3   &:&  \varphi(x) = \frac{1}{2}\sqrt{\xi}
                             (1 + \tanh y \pm  i (1 - \tanh y), 0)
\label{24}
\eea
The energy $E(\varphi)$ of these solutions follows from (\ref{8}):
$E = \frac{4}{3} \xi (2 \l \xi)^{1/2}$ for (\ref{14}), (\ref{17}), (\ref{21}),
whereas $E =  \frac{2}{3} \xi (2 \l \xi)^{1/2}$ for the other ones.

\end{document}